\documentclass[%
 aip,
 amsmath,amssymb,
 reprint,%
]{revtex4-2}

\usepackage{graphicx}% Include figure files
\usepackage{dcolumn}% Align table columns on decimal point
\usepackage{bm}% bold math
\usepackage{xcolor}
\usepackage[normalem]{ulem}
\usepackage[utf8]{inputenc}
\usepackage[T1]{fontenc}
\usepackage{mathptmx}
\usepackage{etoolbox}
\usepackage{color}
\usepackage{hyperref}
\usepackage{amsmath}
\usepackage{cleveref}
\usepackage{CJKutf8}
\usepackage{latexsym}
\usepackage{amstext}
\usepackage{amsfonts}
\usepackage{dsfont}
\usepackage{color}
\usepackage{amssymb}
\usepackage{relsize}
\usepackage{amsxtra}
\usepackage{verbatim}
\usepackage{graphics}
\usepackage{graphicx}
\usepackage{multirow}
\usepackage{ifpdf}
\usepackage{subcaption}
\usepackage{tabularx}
\usepackage{makecell}
\usepackage{ragged2e} % 放在导言区
\makeatletter
\long\def\@makecaption#1#2{%
  \vskip\abovecaptionskip
  {\noindent\parbox{\hsize}{\justifying\textbf{#1.} #2}}%
  \vskip\belowcaptionskip
}
\makeatother
\renewcommand{\thetable}{\arabic{table}}
\makeatletter
\def\@email#1#2{%
 \endgroup
 \patchcmd{\titleblock@produce}
  {\frontmatter@RRAPformat}
  {\frontmatter@RRAPformat{\produce@RRAP{*#1\href{mailto:#2}{#2}}}\frontmatter@RRAPformat}
  {}{}
}%
\makeatother
\begin{document}

\preprint{AIP/123-QED}

\title{Physical Design of Cold Neutron Direct Geometry Inelastic Spectrometer at China Spallation Neutron Source}
% Force line breaks with \\

\author{Qian Zhao}
\affiliation{Institute of High Energy Physics, Chinese Academy of Sciences (CAS), Beijing 100049, China}
\affiliation{Spallation Neutron Source Science Center, Dongguan 523803, China}

\author{Xiaowen Zhang}
\affiliation{Institute of High Energy Physics, Chinese Academy of Sciences (CAS), Beijing 100049, China}
\affiliation{Spallation Neutron Source Science Center, Dongguan 523803, China}

\author{Songwen Xiao}
\affiliation{Institute of High Energy Physics, Chinese Academy of Sciences (CAS), Beijing 100049, China}
\affiliation{Spallation Neutron Source Science Center, Dongguan 523803, China}

\author{Wei Luo}
\affiliation{Institute of High Energy Physics, Chinese Academy of Sciences (CAS), Beijing 100049, China}
\affiliation{Spallation Neutron Source Science Center, Dongguan 523803, China}

\author{Zecong Qin}
\affiliation{Institute of High Energy Physics, Chinese Academy of Sciences (CAS), Beijing 100049, China}
\affiliation{Spallation Neutron Source Science Center, Dongguan 523803, China}

\author{Zhuang Xu}
\affiliation{Institute of High Energy Physics, Chinese Academy of Sciences (CAS), Beijing 100049, China}
\affiliation{Spallation Neutron Source Science Center, Dongguan 523803, China}

\author{Yu Feng}
\altaffiliation{Author to whom correspondence should be addressed. Electronic mail: fengyu@ihep.ac.cn, tongxin@ihep.ac.cn}
\affiliation{Institute of High Energy Physics, Chinese Academy of Sciences (CAS), Beijing 100049, China}
\affiliation{Spallation Neutron Source Science Center, Dongguan 523803, China}

\author{Xin Tong}
\altaffiliation{Author to whom correspondence should be addressed. Electronic mail: fengyu@ihep.ac.cn, tongxin@ihep.ac.cn}
\affiliation{Institute of High Energy Physics, Chinese Academy of Sciences (CAS), Beijing 100049, China}
\affiliation{Spallation Neutron Source Science Center, Dongguan 523803, China}

\begin{abstract}

The Cold-Neutron Inelastic Spectrometer (CNIS) is a direct-geometry, time-of-flight instrument designed for China Spallation Neutron Source (CSNS) and optimized to probe low-energy lattice and magnetic excitations. The instrument integrates a long flight path with bent supermirror guides and an elliptical-focusing geometry to suppress high-energy background while improving cold-neutron delivery to the sample. A flexible multi-disk chopper suite provides pulse shaping, band selection and monochromatization, enabling multi-$E_\textrm{i}$ operation. Modular features, including an interchangeable high-focusing guide insert, radial collimation and a vacuum ``airbox'' for simplified sample-environment integration, enhance signal-to-noise and operational versatility. Through combined flight-path and chopper optimization, CNIS achieves excellent routine-mode energy resolution and can reach approximately $\sim 1\%$ in a dedicated high-resolution configuration. CNIS is planned to commence user operation in 2029, offering a highly flexible platform for cold-neutron inelastic scattering studies.

\end{abstract}

%% Keywords

\maketitle

\begin{figure}[t]
\centering
\includegraphics[width = 0.48\textwidth] {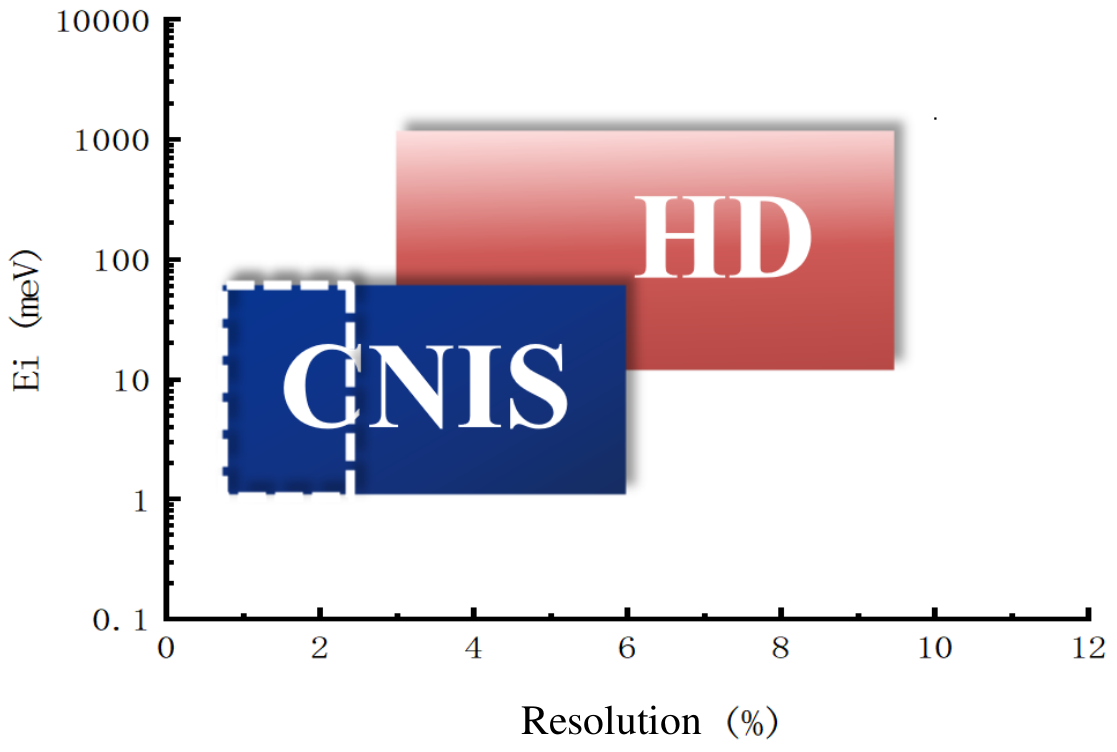}
\caption{
Comparison of energy range and resolution for High-Energy Direct-Geometry Chopper Spectrometer (HD) and Cold-Neutron Inelastic Spectrometer (CNIS). The white dashed rectangle denotes the main target energy and resolution range of the CNIS design. CNIS prioritizes lower incident energies ($E_i$) and delivers finer energy resolution, complementing HD.
}
\label{Compa}
\end{figure}

\begin{figure*}
\centering
\includegraphics[width = 0.92\textwidth] {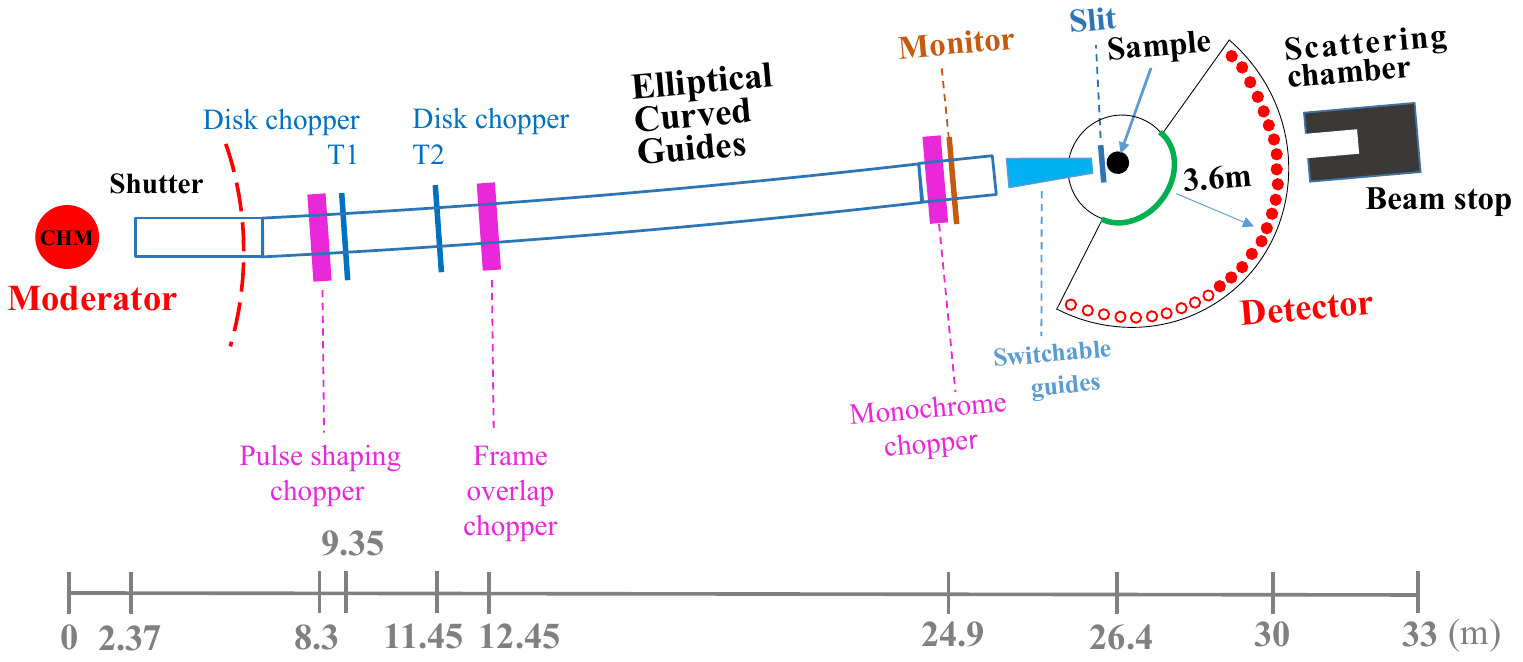}
\caption{
The schematic diagram of CNIS. Locations of various essential components are depicted relative to the surface of the moderator. The schematic is not to scale. From the moderator, a neutron shutter controls the entrance of the neutron beam. The guide and chopper system are designed to direct the desired neutrons to the sample, which will be described in the following sections. The curved section of the neutron guide has a curvature radius of 1500 meters and features an elliptical focusing design in the vertical cross-sectional view. Samples and detectors are installed within a large vacuum chamber, and the beam-stop is designed to absorb the direct beam.
}
\label{CNIS}
\end{figure*}

\begin{figure*}
\centering
\includegraphics[width = 0.92\textwidth] {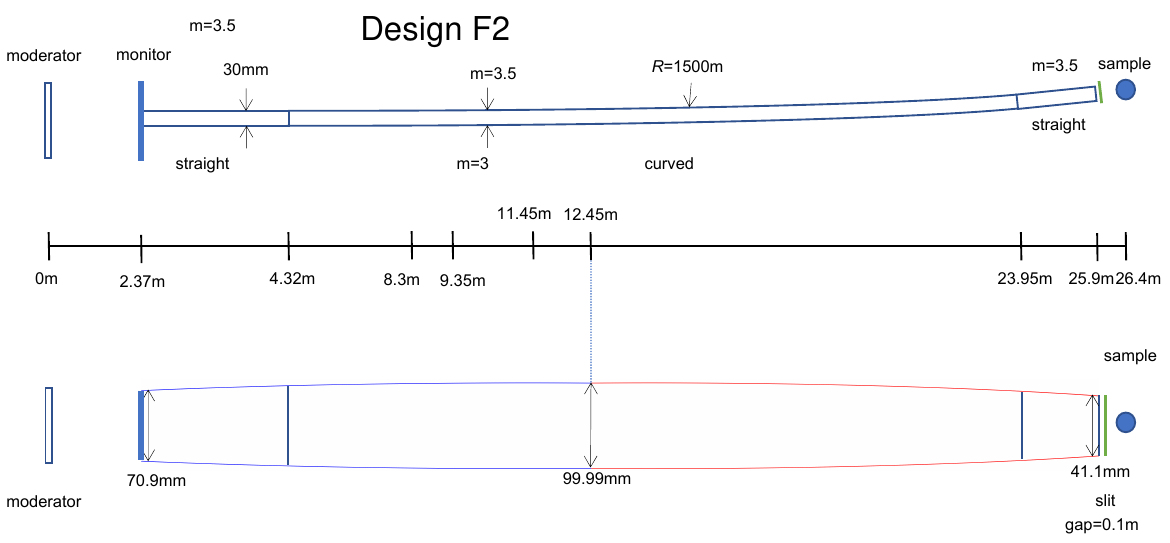}
\caption{
Schematic diagram of the neutron guide tube of the CNIS.  The upper panel shows the horizontal cross-section of the guide: a constant-width (30 mm) straight–bent–straight assembly with a bend radius of 1500 m. The middle panel marks distances from the moderator to some key positions and choppers. The lower panel shows the vertical cross-section, employing a double half-ellipse configuration—-front ellipse in blue and rear ellipse in red—-with the constraint on the maximum guide height indicated.
}
\label{Guide}
\end{figure*}

\section{Introduction}
\label{sec1}
%% Labels are used to cross-reference an item using \ref command.

Inelastic neutron scattering probes energy exchange between neutrons and matter, enabling direct access to lattice and spin dynamics. Beyond static structure, such measurements reveal phonons and magnetic excitations (magnons or spinons) across momentum and energy space, summarized by the dynamic structure factor \(S(Q,\omega)\) with momentum transfer \(Q\) and energy transfer \(\hbar\omega\). These dynamical degrees of freedom are essential to understanding and controlling material properties. \cite{SNSColdNeutron, HODACA2024, CHESS2022, CHESS2018, HuZe2025} Cold neutrons—typically \(0.1\!-\!10\) meV (wavelengths > 3 {\AA})—are well matched to the energy scales of phonons and magnetic excitations, and their de Broglie wavelengths are comparable to lattice spacings. As a result, cold-neutron inelastic scattering provides precise information on vibrational and magnetic modes and generally affords higher energy resolution than hot-neutron instruments, benefiting studies of low-energy excitations in materials such as quantum critical points, ferroelectricity, ferromagnetism, electron–phonon coupling, and magnon–phonon coupling. \cite{unruh2007toftof, bewley2011let, nakajima2011amateras, ollivier2011in5, Deen2021CSPEC}

To meet the evolving needs of diverse experiments and research fields, as a continuation and enhancement of the first phase of the China Spallation Neutron Source (CSNS), the second phase of CSNS was approved at the end of 2023. \cite{chen2016china, WEI200910} CSNS will deliver eleven world-class spectrometers and experimental stations. Within this program, CNIS will be built as the second inelastic instrument, following the High-Energy Direct-Geometry Chopper Spectrometer (HD), providing complementary incident-energy coverage and energy resolution. \cite{LUO2023167676} CNIS is designed to achieve an energy resolution of \(\Delta E/E_i < 2.5\%\) (as low as \(\sim 1\%\) in the high-resolution mode), and is planned for open operation to international researchers in 2029. Figure~\ref{Compa} compares HD and CNIS: CNIS targets lower incident energies (\(E_i\)) and achieves higher energy resolution.

%% Use \subsection commands to start a subsection.

\section{Design of CNIS}

CNIS will be constructed on Beamline 04 of the south hall target station of CSNS, spanning a total length of 33 meters. Figure~\ref{CNIS} illustrates the schematic physical design of CNIS. It primarily comprises the moderator, neutron transport system, chopper system, scattering chamber, and detector system. Neutrons emerge from a coupled liquid-hydrogen moderator, pass through a shutter and travel along a bent, elliptically focused guide that suppresses fast neutrons while choppers reject unwanted wavelength bands. A rapidly rotating monochromating chopper then produces monochromating neutrons—in multi-\(E_i\) mode, three to five incident energies can be delivered. Monitors after the chopper determine the incident energies by time-of-flight, and detectors record scattered neutrons; from the measured time of flight (TOF) changes and hit positions the sample's structural and dynamical properties are obtained.

Energy resolution is the most critical parameter for CNIS, and is determined by several factors—most notably the chopper timing and flight-path geometry, of which \(L_1\), \(L_2\), and \(L_3\) denote the distances from the moderator to the monochromating chopper, from the monochromating chopper to the sample, and from the sample to the detector, respectively. Considering the actual spatial limitations at the construction site, setting \( L_1 = 24.9 \) m, \( L_2 = 1.5 \) m, and \( L_3 = 3.6 \) m is a relatively reasonable choice. Under these conditions, the designed optimal overall energy resolution is 2.5\%.

CNIS employs a neutron-transport strategy that combines bent guides with elliptical focusing and provides two operational modes-high-flux and high-resolution—to meet diverse experimental requirements. The high resolution mode of spectrometer delivers a best-achievable energy resolution of approximately 1\%; other parameters are listed in Table~\ref{GenePara}. The subsequent sections will provide a detailed introduction to the physical design.

\begin{table}
\centering
\renewcommand*{\thetable}{\Roman{table}}
\caption{Characteristic parameters of CNIS}
\label{GenePara}
\begin{tabular} {ll}
\hline
\textbf{Moderator}             & Coupled Liquid Hydrogen
\\
\hline
\textbf{Flight path length}    & $L_{\textrm{moderator-chopper}}$ = 24.9 m
\\
                      &  $L_{\textrm{chopper-sample}}$ = 1.5 m
\\
                      &  $L_{\textrm{sample-detector}}$ = 3.6 m
\\
\hline
\textbf{Incident energy}       & 1-75 meV
\\
\hline
\textbf{Energy resolution}     & $\Delta E$/$E_i$ $\leq$ 2.5\%
\\
\hline
\textbf{Sample Size}& 3 cm $\times$ 3 cm
\\
\hline
\textbf{Scattering angle}      & Horizontal: $\sim$90$^{\circ}$ (reserve the space
\\
                      & for upgrading to \(-28.6^\circ\) to \(140^\circ\).)
\\
                      & Vertical: -18.9$^{\circ}$ -- +27$^{\circ}$
\\
\hline
\textbf{Fast disk choppers}    & 3 (see Table II for details)
\\
\hline
\textbf{Slow disk choppers}    & 2 at $L_{\textrm{moderator-chopper}}$ = 9.35 m \& 11.45 m
\\
\hline
\textbf{Beam transport}        & $m$ = 3 and 3.5 supermirror
\\
                      & Curved section: $L$ = 19.708 m, $R$ = 1500 m
\\
\hline
\textbf{Detector system}       & 208 $^3$He tubes for first term
\\
                      & $\phi$ = 25.4 mm, $L$ = 3,000 mm
\\
\hline
\end{tabular}
\end{table}

\subsection{Moderator}
\label{subsec1}

CNIS employs a Coupled Hydrogen Moderator (CHM), which provides the highest neutron flux intensity within the commonly used incident energy range of 1-20 meV (corresponding to a wavelength range of 9.045–2.02 {\AA}). The moderator is situated beneath the target and has external dimensions of approximately \(10 \times 10 \times 5\) cm\(^3\). The most probable neutron wavelength emitted by the moderator is around 2.8 {\AA}, with a pulse width of approximately 40 $\mu$s/{\AA}. Among the different types of neutron moderators available in CSNS \cite{CHEN2025170431}, the neutron beam from CHM exhibits the highest flux intensity below 20 meV. While CHM delivers high flux at cold energies, its relatively broad time-width requires spectrometer-level optimization (long flight path, chopper shaping) to achieve the desired energy resolution.

%% Use \subsubsection, \paragraph, \subparagraph commands to
%% start 3rd, 4th and 5th level sections.
%% Refer following link for more details.
%% https://en.wikibooks.org/wiki/LaTeX/Document_Structure#Sectioning_commands

\subsection{Neutron Optics}

%\begin{figure}[ht]
  %\centering
  %\begin{subfigure}[b]{0.48\textwidth}
    %\centering
    %\includegraphics[width=\textwidth]{Figure4}
    %\caption{Schematic illustrating the calculation of the neutron direct line of sight (\(L_d\)), denoted by the solid red arrow. \cite{zuo2020bend}}
    %\label{1111}
  %\end{subfigure}
  %\hfill
  %\begin{subfigure}[b]{0.48\textwidth}
    %\centering
    %\includegraphics[width=\textwidth]{Figure5}
    %\caption{Neutron flux before and after passing through the bent guide as simulated by McStas.}
    %\label{fig:subfig2}
  %\end{subfigure}
  %\caption{(a) Schematic illustrating the calculation of the neutron direct line of sight (\(L_d\)), denoted by the solid red arrow. \cite{zuo2020bend} (b) Comparison of neutron wavelength distributions before and after the bent guide, simulated by McStas. The red dotted line represents the neutron distribution at the entrance of the bent guide, while the blue dotted line represents the neutron distribution at the exit of the bent guide.}
  %\label{BentGuide}
%\end{figure}

The neutron guide is a neutron-optical device that transports neutron beams by total external reflection, enabling low-energy neutrons to be conveyed over tens of meters with minimal loss for experiments. In our design, the goals are to maximize cold-neutron flux to improve efficiency and to suppress fast-neutron background to raise the signal-to-noise ratio. Accordingly, we decouple the horizontal and vertical designs: the horizontal cross-section is optimized to reduce fast-neutron background, while the vertical cross-section is optimized to enhance beam intensity. Figure~\ref{Guide} illustrates the schematic diagram of the neutron guides for CNIS with a direct geometry configuration. Horizontally, the guides adopt a straight-bent-straight structure, while vertically, they utilize double semi-elliptical focusing. The neutron guide comprises four segments and the detailed guide parameters are listed in the table~\ref{GuidePara}. Notably, owing to spatial constraints the neutron guide is extended into the scattering chamber. This yields two advantages: it permits a longer guide within the confined volume, thereby improving neutron flux; and it removes the aluminum window between the guide and chamber, reducing experimental background.

In CNIS, bent neutron guides are employed to effectively suppress background from fast neutrons. We note that some cold-chopper spectrometers achieve very low background with a straight primary guide combined with carefully phased choppers and shielding (e.g., the LET instrument at ISIS).\cite{bewley2011let} For the present instrument, however, the combination of our moderator geometry, desired instrument layout and local shielding strategy led us to adopt a curved section in the guides configuration. The bender removes direct line-of-sight from the moderator and thereby reduces fast-neutron and prompt-$\gamma$ leakage that could otherwise reach the sample/detector outside the cold-neutron measurement window. The selection of the curvature radius \( R \) for the neutron bent guide requires a comprehensive consideration of multiple factors. If the curvature radius is too large, it can result in a direct line of sight for neutrons, allowing high-energy neutrons to pass through the bent guide without being intercepted. Conversely, if the curvature radius is too small, it may lead to significant loss of neutron beam intensity and an excessively large cutoff wavelength, adversely affecting the practical application of the cold neutron spectrometer. Since the guide’s cutoff wavelength obeys \(\lambda=\sqrt{2d/(nR)}/(m\theta_c)\), a smaller \(m\) overly suppresses high-\(E\) neutrons while a larger \(m\) drives up cost; thus, with \(d=3\) cm, \(n=1\), \(R=1500\) m and \(\theta_c=0.00173\) rad we adopt \(m=3.5\) (yielding \(\lambda\approx1.04\,\text{\AA}\)) and set the inner surface to \(m=3.0\) to control cost without sacrificing beam intensity.

\begin{table}
\centering
\renewcommand*{\thetable}{\Roman{table}}
\caption{Designed Parameters of Neutron Guides}
\label{GuidePara}
\begin{tabular} {llll}
\hline
              & \textbf{Shape}                                        & \textbf{$m$-value}            & \textbf{Position} (m)
\\
              &                                              &                      & (from moderator)
\\
\hline
  Section 1   & Straight guide          & 3.5                                       & 2.37–4.32
\\
              & divergent ellipse            &                                        &
\\
\hline
  Section 2   & Bent guide, $R$ = 1500 m     & inner: 3.0;   & 4.32–12.45
\\
              & divergent ellipse            & other: 3.5                                       &
\\
\hline
  Section 3   & Bent guide, $R$ = 1500 m     & inner: 3.0;   & 12.45–23.95
\\
              & convergent ellipse           & other: 3.5                                        &
\\
\hline
  Section 4   & Straight guide           & 3.5                                       & 23.95–25.9
\\
              & convergent ellipse            &                                        &
\\
\hline
\end{tabular}
\end{table}

\begin{figure}[!t]
\centering
\includegraphics[width = 0.48\textwidth] {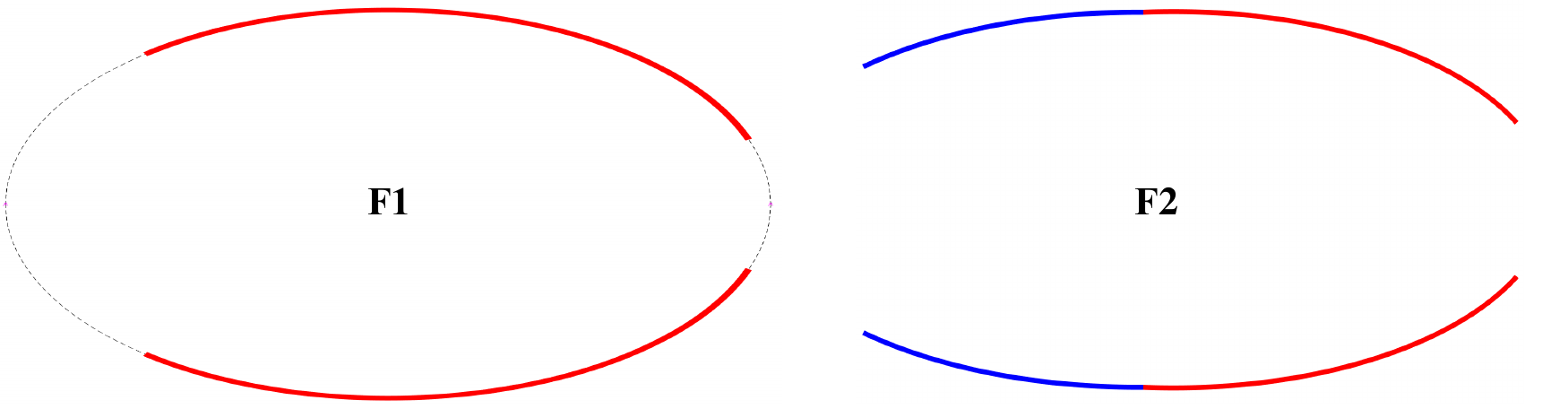}
\caption{
Two candidate elliptical neutron guide configurations.
}
\label{2Guides}
\end{figure}

In the vertical cross-sectional view, we adopt elliptical focusing neutron guides to enhance neutron flux. To maximize the cold neutron beam, McStas and Optuna Optimization Framework \cite{akiba2019optuna} were utilized to optimize all parameters.\cite{zhou2025bayesian} Two elliptical-guide layouts were evaluated(Fig.~\ref{2Guides}). We first adopted a full-elliptical design (F1): within a certain range, increasing the guide entrance and exit apertures increases the neutron flux, which in turn pushed the vertical height at the ellipse center toward the chopper clearance limit (100 mm). Under this constraint, we applied the Optuna algorithm to automatically optimize the ellipse’s front and rear focal points. Because the full elliptical design exhibits discontinuities in the angular-divergence dependence of wavelength, we then moved to a decoupled alternative (F2)—a pair of half ellipses \cite{sala2022chess}—with a height constraint at the 12.45 m chopper. In this design, height constraints are applied at the 12.45 m chopper position, with the rear focus of the front ellipse fixed at the sample position and the front focus of the rear ellipse fixed at the moderator position. The Optuna algorithm \cite{akiba2019optuna} is used to automatically optimize the positions of the front focal point of the front ellipse and the rear focal point of the rear ellipse. After Optuna optimization, the double half-elliptical combination (F2) achieved comparable or higher neutron beam flux and provided a smoother divergence distribution at the sample.

\begin{figure}
\centering
\includegraphics[width = 0.42\textwidth] {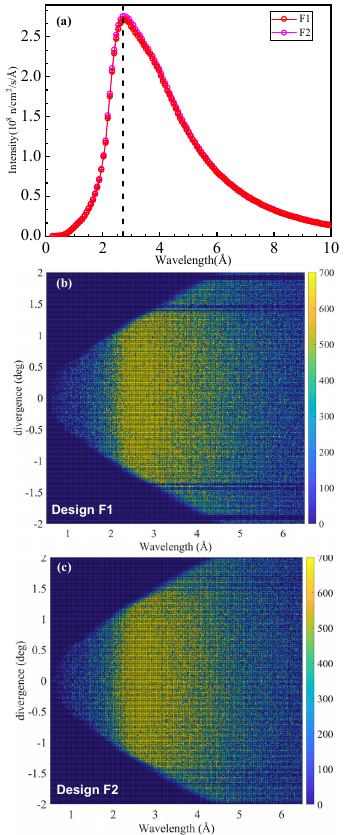}
\caption{
(a) Simulated neutron beam intensity wavelength distribution at the sample position, with the black dotted line indicating the wavelength position where the beam is strongest; (b) (c) Neutron beam vertical divergence angle wavelength distribution at the sample position from simulation.
}
\label{IntenWav}
\end{figure}

\begin{figure*}[t]
\centering
\includegraphics[width = 0.95\textwidth] {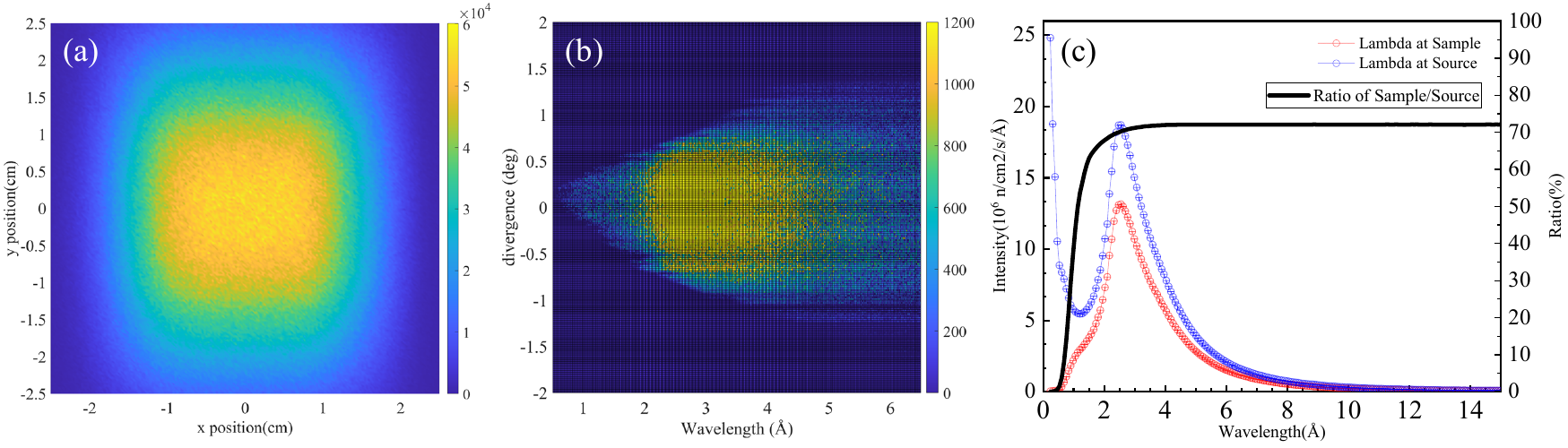}
\caption{
(a) Simulated neutron beam spot within a 5 $\times$ 5 cm$^2$ area at the sample position; (b) Simulation results of the horizontal divergence angle at the sample position as a function of wavelength. (c) Black solid line is the simulated Brilliance Transfer factor. Blue/red void circles are wavelength distribution of neutron flux at source/sample.
}
\label{Spot}
\end{figure*}

Figure~\ref{IntenWav}(a) presents the wavelength distribution of neutron beam flux at the sample position as simulated by McStas. It can be observed that both the full elliptical design and the double half-elliptical design exhibit similar performances. The designs achieve their maximum neutron beam flux intensity at 2.7 {\AA} (11.2 meV), and the distribution aligns with the experimental requirements of the cold neutron inelastic spectrometer.

At the same time, we simulated the wavelength distribution of neutron divergence angles at the sample position as shown in fig.~\ref{IntenWav}(b)(c). It can be observed that Design F1, the full elliptical design, exhibits discontinuities in divergence at different energies, whereas the double half-elliptical combination design maintains continuous divergence at the sample position, meaning that the neutron divergence arriving at the sample under monochromating mode is uniform, thereby outperforming the full elliptical design. We calculated the neutron beam spot at the sample position for the optimized double half-elliptical guide configuration using McStas (see fig.~\ref{Spot}(a)). Under full-band (white-beam) conditions, within a 5 $\times$ 5 cm$^2$ area at the sample position the neutrons are predominantly concentrated within the target 3 $\times$ 3 cm$^2$ region with high intensity, satisfying the design requirements of the cold-neutron inelastic spectrometer. The horizontal divergence at the sample position has also been evaluated by simulation, as shown in fig.~\ref{Spot}(b). The results indicate that, with a curved guide employed in the beamline, the horizontal divergence remains well controlled within the designed acceptance range, demonstrating that the bender does not introduce excessive phase-space broadening in the horizontal direction. We have also calculated the Brilliance Transfer (BT) as a function of wavelength (fig.~\ref{Spot}(c)) which evaluates the performance of the neutron guide by measuring the ratio of neutrons reaching the sample within a divergence angle range of $\pm$0.5$^\circ$ and a sample area of 3 cm $\times$ 3 cm relative to those emitted by the moderator. The simulation results show a BT of approximately 70\% for wavelengths above 3 {\AA}, demonstrating good neutron transport efficiency. The reduction from the ideal value mainly arises from unavoidable installation gaps between guide sections, the use of a curved guide, and the distance between the guide exit and the sample position. In summary, we finally selected the Double Half-Elliptical Combination Design as the final guide configuration design.

In addition, we have added a detachable, interchangeable insert to the end of the guide that can be configured as a collimating guide or as a flight tube; in particular, a high-focusing guide ($m$ = 6). This 500 mm-long section, with an exit aperture of 10 mm $\times$ 10 mm, is specifically designed for smaller samples and can increase the neutron flux within a 1 cm$^2$ area to more than 2.5 times its original value. Under the 100 kW source operation mode, the simulated neutron flux at the sample position is 4.196 × 10$^8$ n/cm$^2$/s averaged over a 3 $\times$ 3 cm$^2$ area, and it increases to 1.0862 × 10$^9$ n/cm$^2$/s when the $m$ = 6 high-focusing guide is installed.

\subsection{Choppers}

\begin{table}[b]
  \centering
  \renewcommand*{\thetable}{\Roman{table}}
  \caption{Technical parameters of the CNIS choppers}
  \label{Chopper}
  \begin{tabular}{llc}
    \hline
    \textbf{Chopper} & \textbf{Technical Parameter} & \textbf{Value} \\
    \hline
    Double-disk Pulse Shaping & Operating Speed & 300 Hz \\
    Chopper P & Installation Position & 8.3 m \\
    \hline
    Double-disk Bandwidth & Operating Speed & 12.5/25 Hz \\
    Chopper T1 & Installation Position & 9.35 m \\
    \hline
    Double-disk Bandwidth & Operating Speed & 12.5/25 Hz \\
    Chopper T2 & Installation Position & 11.45 m \\
    \hline
    Single-disk Fast & Operating Speed & 300 Hz \\
    Chopper R & Installation Position & 12.45 m \\
    \hline
    Double-disk Monochromating & Operating Speed & 300 Hz \\
    Chopper M & Installation Position & 24.9 m\\
    \hline
  \end{tabular}
\end{table}

CNIS is a cold neutron inelastic spectrometer that fully utilizes the flight time characteristics of neutron pulses from a pulsed spallation source. Due to its design, which employs bent guides to eliminate background noise from fast neutrons and $\gamma$-rays, CNIS does not require the installation of T0 choppers used to block high-energy neutrons and $\gamma$-rays. Instead, it can rely solely on bandwidth choppers to remove long-wavelength backgrounds and select the incident energy range. CNIS will be equipped with two slow bandwidth choppers, offering a maximum transmitted open bandwidth of 8 {\AA}; phase adjustment enables an effective opening range of 0$^{\circ}$–175$^{\circ}$, accommodating multiple experimental modes. Our design draws inspiration from the AMATERAS cold-neutron inelastic spectrometer at J-PARC \cite{nakajima2011amateras}.

\begin{figure}[b]
\centering
\includegraphics[width = 0.48\textwidth] {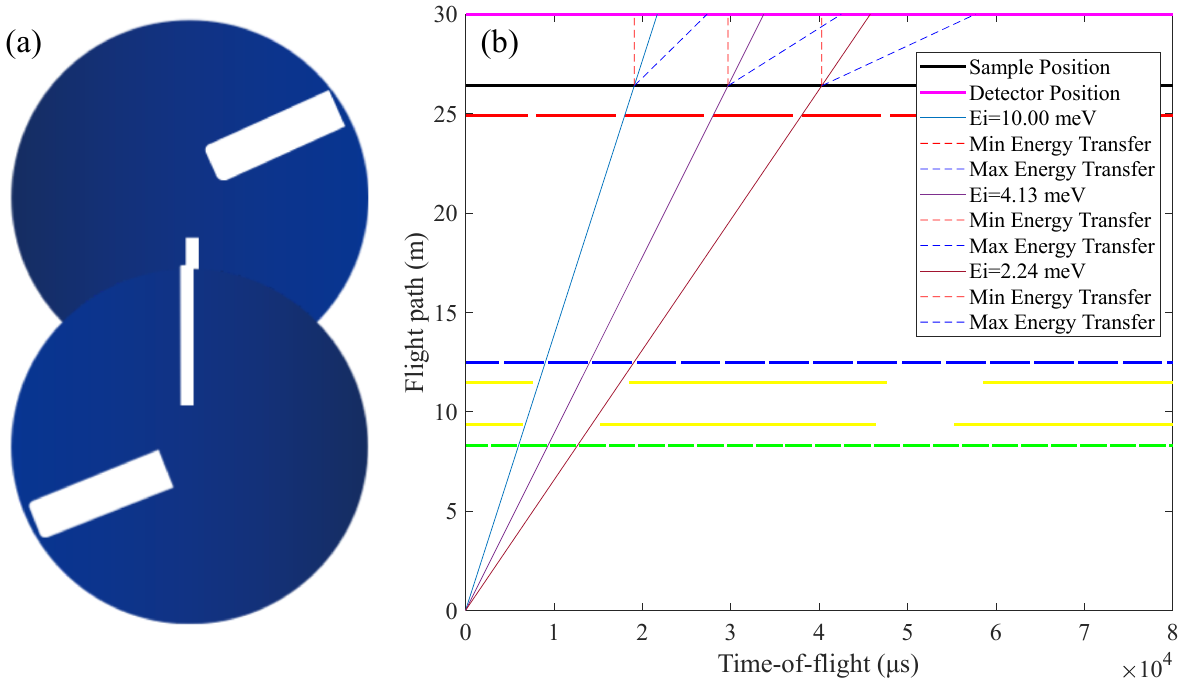}
\caption{
(a) Schematic of the vertically oriented double-disk chopper. (b) Calculated time-distance diagram in the multi-Ei mode, showing the operating frequencies and phases of the choppers.
}
\label{PChopper}
\end{figure}

For a direct geometry spectrometer, it is required that the neutron beam incident on the sample has a single known energy. This allows for the calculation of the flight time of neutrons undergoing inelastic scattering at the sample, thereby determining the energy changes during the inelastic scattering process. Consequently, monochromatization of the neutron beam is necessary. CNIS will be equipped with three pulse choppers: Pulse Shaping Chopper (P chopper), Wavelength Removing Chopper (R chopper), and Monochromating Chopper (M chopper). Their rotation speed can reach 300 Hz with a 30 ns time window. Their functions are to adjust the time width of neutrons exiting the moderator, remove unwanted incident energies and long-wavelength tail effects, and monochromatize the incident neutrons, respectively. Notably, the spacing of the P, R and M choppers is chosen in the ratio 2:3:6. This arrangement enables a multi-$E_\textrm{i}$ mode: with a single chopper configuration multiple desired incident energies are produced simultaneously, thereby improving experimental efficiency.

On the monochromating chopper’s two disks, two opposing slits of different widths (30 mm and 10 mm) are incorporated; reducing the slit from 30 mm to 10 mm decreases the neutron flux while narrowing the incident peak FWHM, thereby enabling high-flux mode and high-resolution mode. Importantly, in our design the P chopper--like the M chopper--employs a vertically oriented, counter-rotating double-disk configuration (as shown in the fig.~\ref{PChopper}(a)). This configuration provides higher neutron transmission than a parallel-mounted double-disk chopper, thereby increasing flux at a given resolution. The specific parameters of the choppers are presented in Table~\ref{Chopper}. Kenji Nakajima et al. \cite{nakajima2011amateras} proposed a simple estimation of the energy resolution for a chopper spectrometer equipped with a pulse-shaping chopper. Using this expression with the CNIS design parameters, we find a fractional energy resolution \(\Delta E/E_i < 2.5\%\); in the high-resolution mode, the top energy resolution can reach \(\sim 1\%\). A simulated time--distance diagram for the multi-$E_i$ chopper system (e.g., $E_i = 10$~meV) is shown in fig.~\ref{PChopper}(b), with high-speed choppers operating at 300, 200, and 100 Hz, and a bandwidth chopper at 25 Hz. The simulated neutron flux at the sample ($3\,\mathrm{cm}\times3\,\mathrm{cm}$, $m = 3/3.5$, no slit) is $0.748$, $3.879$, $3.329$, and $0.562 \times 10^6~\mathrm{n/cm^2/s}$ for $E_i = 1$, $10$, $11.2$, and $20$~meV, respectively; these values will be further validated during commissioning.

\subsection{Scattering Chamber and Radial Collimator}

\begin{figure}
\centering
\includegraphics[width = 0.48\textwidth] {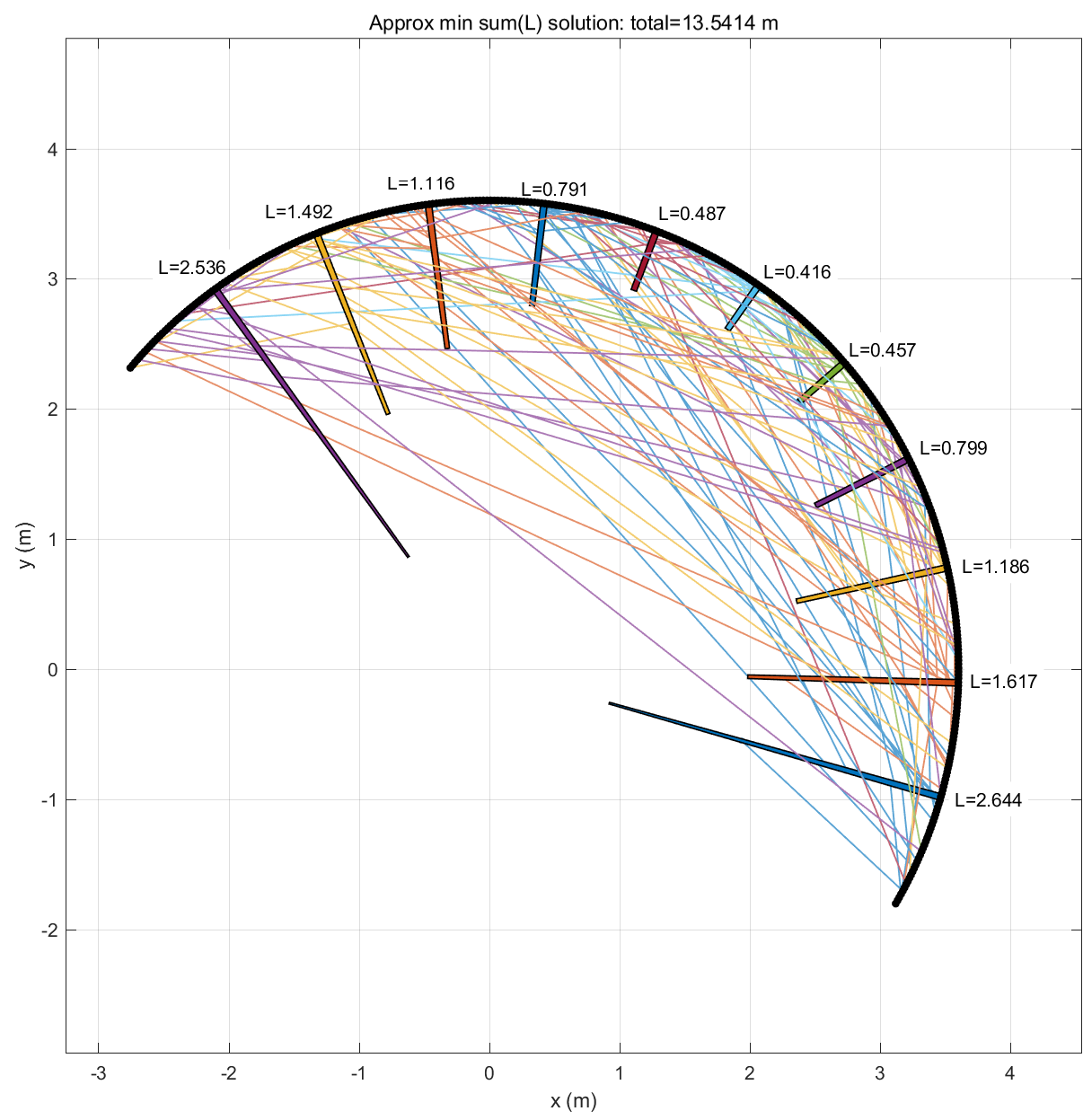}
\caption{
Schematic and computed design length of isolation vanes between detector banks for cross-talk suppression; $^3$He tubes are abstracted into black pixels and pairwise lines are used to determine the minimal total vane length.
}
\label{Vanes}
\end{figure}

\begin{figure}
\centering
\includegraphics[width = 0.42\textwidth] {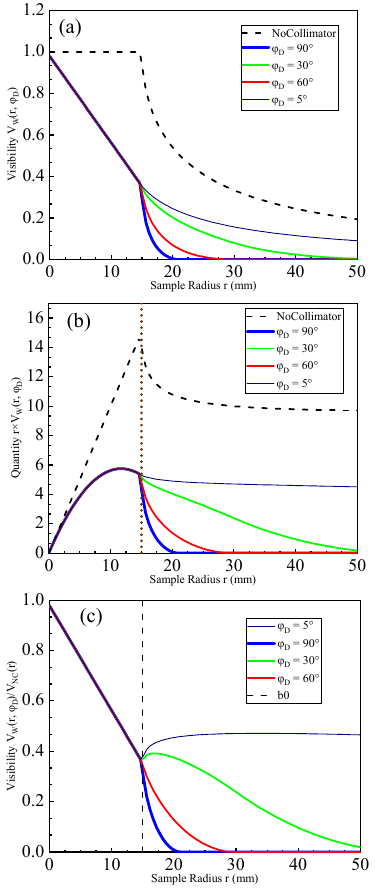}
\caption{
(a) Visibility $V_W(r,\phi_D)$ for several scattering angles $\phi_D$. (b) Corresponding scattered intensity; the vertical dashed line indicates the sample radius. (c) Visibility ratio $V_W/V_{NC}$, highlighting the reduction of off-sample contributions provided by the radial collimator.
}
\label{Colli}
\end{figure}

The vacuum scattering chamber forms the core of the cold-neutron direct-geometry inelastic spectrometer. It provides both the structural support and installation space for the detectors, sample, and sample environments, while ensuring a high-vacuum, low-background environment for neutron scattering. A key aspect of the chamber design lies in its connection with the detector system. To address these constraints, we adopted an "airbox" solution, in which all electronics are encapsulated in a dedicated enclosure. This approach significantly relaxes the chamber’s vacuum specifications, reduces machining and deformation requirements, and simplifies maintenance.

The spectrometer is designed to accommodate standard large sample environments, including superconducting magnets up to 14 T, with sufficient space and angular coverage ensured by the sample position, detector geometry, and collimator layout. Non-magnetic materials are used near the sample and scattering chamber to minimize magnetic interference, and the sample chamber bottom is made of aluminum based on stray-field considerations.

For neutron detection, most established spectrometers worldwide employ either high-pressure \(^3\)He tubes or multi-wire proportional chambers (MWPCs). \cite{unruh2007toftof, bewley2011let, nakajima2011amateras, ollivier2011in5, Deen2021CSPEC, kajimoto20134SEASONS, faak2022panther, bewley2006merlin} Compared with high-pressure \(^3\)He-MWPCs, two-dimensional detectors based on arrays of \(^3\)He tubes offer clear advantages in installation and maintenance: individual tubes can be replaced rapidly, avoiding the risk of total detector failure caused by a single broken wire in an MWPC. Based on these considerations, the CNIS has chosen a large-area \(^3\)He tube array at a pressure of 10 atm as its primary detector system.

The detector requirements are defined by the spectrometer geometry. With a sample-to-detector distance of 3.6 m and a beam height of 1661 mm, a 3 meter detector provides vertical coverage from -18.9$^{\circ}$ to +27$^{\circ}$. In the horizontal direction, the detectors configuration covering a scattering angle of $\sim$90$^{\circ}$. Moreover, there remains capacity for future upgrades to 170$^{\circ}$ coverage. In summary, the decision to adopt \(^3\)He tube detectors reflects a balance of multidisciplinary experimental needs, international best practice, beam power considerations, and upgrade potential, ensuring both present capability and long-term adaptability of the CNIS.

The Q resolution is primarily governed by the moderator pulse width and the angular divergence, which depends on detector geometry, sample size, and crystal quality. It improves with increasing scattering angle, reaching better than 10\% above 20$^{\circ}$ and approximately 1\% above 100$^{\circ}$.

Between detector banks we have implemented isolation vanes to suppress cross-talk between adjacent detectors. Each vane is approximately one detector tube width thick (25.4 mm). The required vane length was obtained by a simple geometric procedure: the deployed $^3$He tubes were abstracted into 384 pixels, pairwise lines were drawn between pixels, and an algorithm computed the minimal total vane length needed to block those lines; the resulting design length is shown in the fig.~\ref{Vanes}. The presented results should be regarded as preliminary and intended for initial design guidance only. Practical installation must still take on-site space constraints into account.

The radial collimator is installed between the sample environment and the detector array and is formed by a series of radially arranged neutron-absorbing vanes. The gaps between adjacent vanes define the neutron flight paths. By intercepting stray neutrons scattered from the sample environment and ancillary equipment, the radial collimator substantially improves the spectrometer signal-to-noise ratio.

The design of radial collimator was validated by numerical simulations (MATLAB) that evaluate visibility as a function of neutron incident angle and sample size, and the results were depicted in fig.~\ref{Colli}. \cite{copley1994analysis} This representation makes it straightforward to compare how much of the measured signal originates from the sample region versus off-sample contributions and to quantify the effect of collimation or other geometry-dependent weighting functions on the integrated scattering. \cite{copley1994analysis} In the analysis we also compared the visibility with the radial collimator, $V_W(r,\phi_D)$, to that without a collimator, $V_{NC}(r)$. Simulation results show that the radial collimator effectively removes high-angle neutrons originating outside the sample, thereby enhancing visibility and improving overall data quality.

\section{Summary}
\label{app1}

In summary, the CNIS instrument uses a long flight path combined with a curved, elliptically focused neutron guide and a sophisticated multi-chopper system to achieve high neutron flux and good energy resolution (optimally 1\%). A detachable insert is introduced at the end of the guide, including high-focusing guide ($m$ = 6) which can increase the neutron flux for small samples by more than 250\%. CNIS is scheduled to begin for opening user operation in 2029 and, together with HD, will substantially extend research capabilities across a broader energy range of CSNS.

%Appendix text.

%% For citations use:
%%       \cite{<label>} ==> [1]

%%
%Example citation, See \cite{lamport94}.

%% If you have bib database file and want bibtex to generate the
%% bibitems, please use
%%
%%  \bibliographystyle{elsarticle-num}
%%  \bibliography{<your bibdatabase>}

%% else use the following coding to input the bibitems directly in the
%% TeX file.

%% Refer following link for more details about bibliography and citations.
%% https://en.wikibooks.org/wiki/LaTeX/Bibliography_Management

%\begin{thebibliography}{00}

%% For numbered reference style
%% \bibitem{label}
%% Text of bibliographic item

%\bibitem{lamport94}
%  Leslie Lamport,
%  \textit{\LaTeX: a document preparation system},
%  Addison Wesley, Massachusetts,
%  2nd edition,
%  1994.

%\end{thebibliography}

\section{Acknowledgement}

This work is part of the second phase of China Spallation Neutron Source. Work of Y. Feng is supported by National Key Research and Development Program of China (Grant No. 2022YFA1604104) and the Large Scientific Facility Open Subject of Songshan Lake, Dongguan, Guangdong (Grant No. KFKT2022A03). Work of X. Tong is supported by the National Science Fund for Distinguished Young Scholars (Grant No. 12425512).

\bibliographystyle{elsarticle-num}
\bibliography{CNIS}

\end{document}